\documentclass{article}
\usepackage{spconf,amsmath,graphicx}
\usepackage{booktabs}
\usepackage{makecell}
\usepackage{multirow}
\usepackage{tikz}
\usepackage{url}
\usepackage{amssymb}
\usepackage{pifont}
\newcommand{\cmark}{\ding{51}}%
\newcommand{\xmark}{\ding{55}}%

\usepackage{tabu}
\usepackage{tabularx} 

\usepackage{subfig}

\usepackage{listings}
\usepackage{color}

\usepackage{acronym}
\usepackage[font=small,skip=0pt]{caption}

\definecolor{dkgreen}{rgb}{0,0.6,0}
\definecolor{gray}{rgb}{0.5,0.5,0.5}
\definecolor{mauve}{rgb}{0.58,0,0.82}

\makeatletter
\newcommand{\ssymbol}[1]{^{\@fnsymbol{#1}}}
\makeatother

\title{Multi-turn RNN-T for streaming recognition of multi-party speech}
\name{Ilya Sklyar$^{1*}$, Anna Piunova$^{1*}$, Xianrui Zheng$^{2+}$, Yulan Liu$\ssymbol{2}$\thanks{$^*$These authors have contributed equally. $^{+}$The contribution was fully conducted during internship in Amazon. $\ssymbol{2}$Work was done while in Amazon.}}
\address{$^1$Amazon Alexa, $^2$University of Cambridge\\
	\small \ttfamily ilsklyar@amazon.com\hspace{1cm}piunova@amazon.com\hspace{1cm}xz396@eng.cam.ac.uk
}
\begin{document}
\ninept
\maketitle
%

\newacro{AIR}[AIR]{acoustic impulse response}
\newacro{ASR}[ASR]{Automatic speech recognition}
\newacro{cpWER}[cpWER]{concatenated minimum-permutation word error rate}
\newacro{CTC}[CTC]{connectionist temporal classification}
\newacro{DAT}[DAT]{deterministic assignment training}
\newacro{DAT-MS-RNN-T}[DAT-MS-RNN-T]{MS-RNN-T with deterministic assignment training}
\newacro{E2E}[E2E]{end-to-end}
\newacro{ISM}[ISM]{image source method}
\newacro{LSTM}[LSTM]{long short term memory}
\newacro{MS-RNN-T}[MS-RNN-T]{multi-speaker recurrent neural network transducer}
\newacro{MT-MS-RNN-T}[MT-MS-RNN-T]{multi-turn multi-speaker RNN-T}
\newacro{MT-RNN-T}[MT-RNN-T]{multi-turn RNN-T}
\newacro{OB}[OB]{overlap-based}
\newacro{OED WER}[OED WER]{optimal edit distance WER}
\newacro{ORC}[ORC]{optimal reference combination}
\newacro{ORC WER}[ORC WER]{optimal reference combination WER}
\newacro{PIT}[PIT]{permutation invariant training}
\newacro{PIT-MS-RNN-T}[PIT-MS-RNN-T]{MS-RNN-T with permutation invariant training}
\newacro{RIRs}[RIRs]{room impulse responses}
\newacro{RNN-T}[RNN-T]{recurrent neural network transducer}
\newacro{SB}[SB]{speaker-based}
\newacro{SNR}[SNR]{speech-to-noise ratio}
\newacro{SOT}[SOT]{serialized output training}
\newacro{WER}[WER]{word error rate}
\newacro{SURT}[SURT]{streaming unmixing and recognition transducer}

\newacroindefinite{ASR}{an}{an}
\newacroindefinite{E2E}{an}{an}
\newacroindefinite{RNN-T}{an}{an}

\begin{abstract}
\ac{ASR} of single channel far-field recordings with an unknown number of speakers is traditionally tackled by cascaded modules.
Recent research shows that \ac{E2E} multi-speaker \ac{ASR} models can achieve superior recognition accuracy compared to modular systems.
However, these models do not ensure real-time applicability due to their dependency on full audio context.
This work takes real-time applicability as the first priority in model design and addresses a few challenges in previous work on \ac{MS-RNN-T}.
First, we introduce on-the-fly overlapping speech simulation during training, yielding 14\% relative \ac{WER} improvement on LibriSpeechMix test set.
Second, we propose a novel \ac{MT-RNN-T} model with an overlap-based target arrangement strategy that generalizes to an arbitrary number of speakers without changes in the model architecture. 
We investigate the impact of the maximum number of speakers seen during training on MT-RNN-T performance on LibriCSS test set, and report 28\% relative WER improvement over the two-speaker \ac{MS-RNN-T}.
Third, we experiment with a rich transcription strategy for joint recognition and segmentation of multi-party speech. Through an in-depth analysis, we discuss potential pitfalls of the proposed system as well as promising future research directions.
\end{abstract}

\section{Introduction}
\label{sec:introduction}

Recognizing multi-party speech from an arbitrary number of speakers with naturally occurring speech overlap in a far-field environment is a challenging research problem that has been extensively studied for years \cite{Hain2007_AMI,fox13b_interspeech,Liu2016_SWC2,Barker_2018}.
This problem is frequently referred to as the ``cocktail party problem''.
Its solution holds the key to the next era of speech technology, and lays the foundation for natural human-device voice interactions.
In such interactions, the speech agent on smart device is expected to navigate through multi-party conversations and provide the support in need.
To allow real-time human-device interactions, the \ac{ASR} system needs to provide transcripts in a streaming fashion for all speakers in the conversation.
For better privacy protection, the \ac{ASR} system design needs to accommodate local operation on device only, which means the model should be light-weight and robust against acoustic front-end limitations.
All these factors point to the importance of streaming speech recognition of an arbitrary number of speakers with potential speech overlaps using single distant microphone only.

Traditionally, multi-speaker \ac{ASR} was approached with cascaded modules such as speech separation and speech recognition \cite{Isik_2016, Menne_2019, Settle_2018, von_Neumann_2020, Neumann_2020}.
Due to the system complexity and technical limitation in some modules, such systems are more suitable for off-line applications where the full audio recording is available before transcription task starts.
In past years, research progress has been made in training joint models that optimize multi-speaker ASR performance directly \cite{Yu_2017_icassp, Yu_2017_interspeech, Qian_2018, Seki_2018, Chang_2019, Tripathi_2020}.
Recently, \cite{k2020serialized} proposed joint multi-speaker speech recognition and speaker change detection for any number of speakers via \ac{SOT}, and demonstrated its effectiveness on simulated multi-speaker test set of LibriSpeechMix.
Later \cite{kanda2021largescale} showed that \ac{SOT} is also effective on real multi-speaker meeting corpus, and subsequent works \cite{kanda2020joint, kanda2021slt, kanda2021endtoend, kanda2021comparative} extended \ac{SOT} into speaker-attributed ASR that can transcribe ``who spoke what" in multi-speaker conversations with one integrated model.

In parallel, recent research also started to look at maintaining the same performance under streaming conditions.
Two conceptually similar streaming multi-speaker ASR models based on a \ac{RNN-T} \cite{graves2012sequence}, i.e. \ac{MS-RNN-T} \cite{SklyarPiunovaLiu_streaming_rnnt_icassp2021} and \ac{SURT} \cite{Lu2021}, were concurrently proposed to allow time-synchronous decoding of partially overlapping speech, and \cite{Lu2021} was later extended with speaker identification in \cite{lu2021streaming}.
\cite{SklyarPiunovaLiu_streaming_rnnt_icassp2021} conducted evaluations of \ac{MS-RNN-T} on single-speaker LibriSpeech and two-speaker LibriSpeechMix.
It showed that \ac{MS-RNN-T} achieves on-par performance with off-line multi-speaker ASR systems on two-speaker test data and also improves robustness of the system on noisy single-speaker test-other partition of LibriSpeech.

This work builds on \cite{SklyarPiunovaLiu_streaming_rnnt_icassp2021}, with the following contributions.
First, we show the advantage of on-the-fly overlapping speech simulation over fixed pre-simulated data for training, leading to a new low \ac{WER} on two-speaker partition of LibriSpeechMix with a streamable model architecture.
Second, we address the main limitation of \ac{MS-RNN-T}, i.e. the hard restriction on the maximum number of speakers in the audio.
Assuming there are up to two speakers overlapping at a time, the proposed \ac{MT-RNN-T} model is capable to recognize speech from an arbitrary number of speakers.
Additionally, we make the first step towards streaming rich transcription of multi-party speech by introducing segmentation tag.
Last but not least, we analyze performance in-depth using proposed evaluation metrics, \ac{ORC} WER and turn counting accuracy, and discuss potential future work.

\section{Prior work}
\label{sec:prior-work}

\begin{figure}[ht]
	\includegraphics[width=\linewidth]{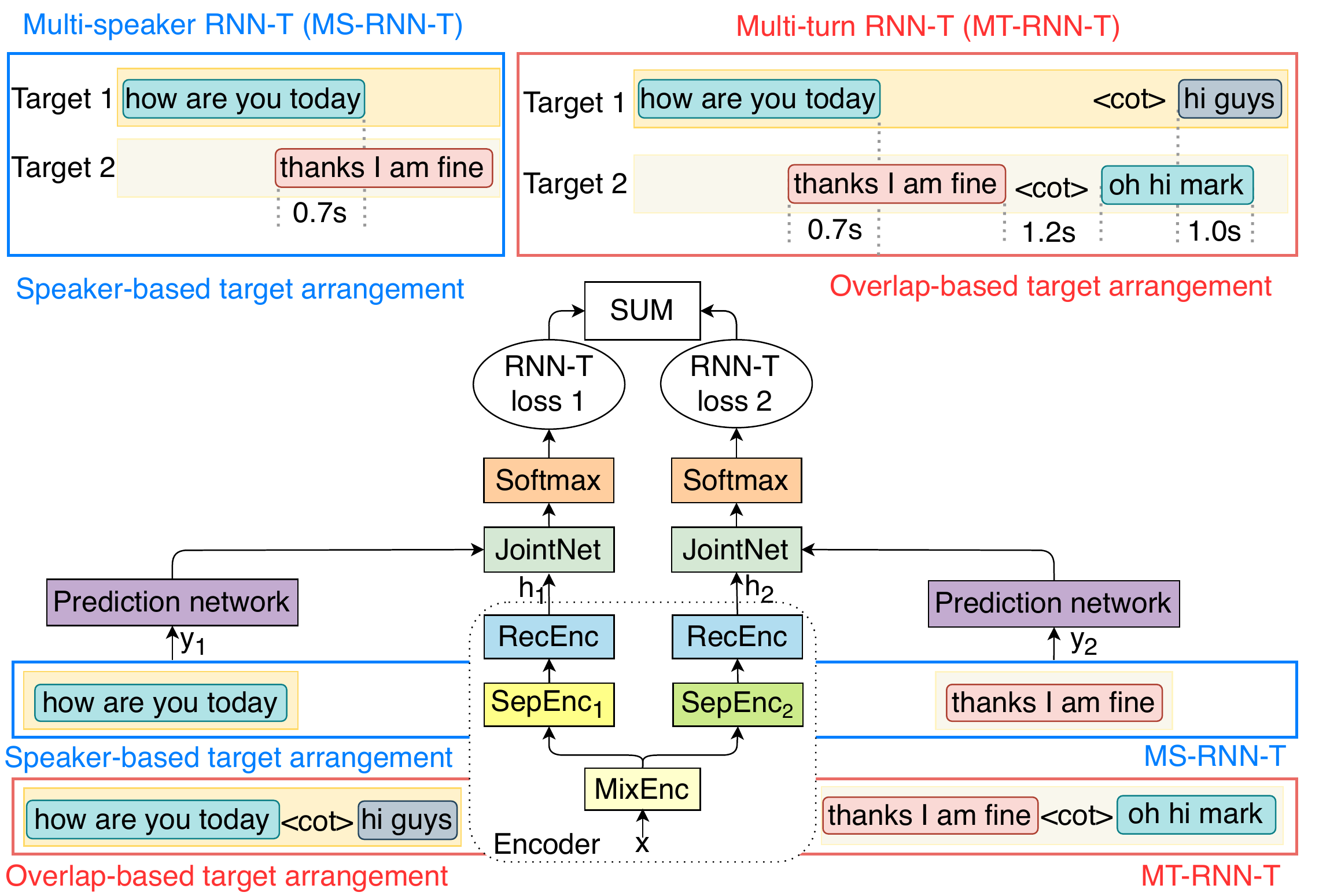}
	\vspace*{0.1mm}
	\caption{Baseline \ac{MS-RNN-T} with speaker-based target arrangement from \cite{SklyarPiunovaLiu_streaming_rnnt_icassp2021}, and proposed \ac{MT-RNN-T} with overlap-based target arrangement with optional \texttt{<cot>} tag for speech segmentation. Model blocks with the same colour have tied parameters, transcripts in the colour-matched boxes belong to the same speaker.}
	\label{fig:mt-ms-rnnt}
\end{figure}


\ac{MS-RNN-T} \cite{SklyarPiunovaLiu_streaming_rnnt_icassp2021} extends the standard \ac{RNN-T} \cite{graves2012sequence} to overlapping speech recognition with multiple output channels \(N = S\), where $S$ is the number of speakers in the audio.
The encoder of \ac{MS-RNN-T} has a modular structure containing a mixture encoder (\(\operatorname{MixEnc}\)), $N$ separation encoders (\(\operatorname{SepEnc}_n\)) for each output channel \(n \in \{1,...,N\} \) and a recognition encoder (\(\operatorname{RecEnc}\)) with shared parameters between output channels. 
The encoder takes acoustic features \(\mathbf{x}\) as input and produces high-level disentangled acoustic representations \( \mathbf{h}_n\) as output, as described mathematically in Eq. \ref{eq:speech-encoder} and visualized in the corresponding block in Fig. \ref{fig:mt-ms-rnnt}.
\vspace{-0.1cm}
\begin{align}
\label{eq:speech-encoder}
\mathbf{h}_n = \operatorname{RecEnc}(\operatorname{SepEnc}_n(\operatorname{MixEnc}(\mathbf{x}))) 
\end{align}
\cite{SklyarPiunovaLiu_streaming_rnnt_icassp2021} proposed two ways to associate \(\mathbf{h}_n \) with prediction network outputs for each label sequence \(\mathbf{y}_{n}\): \ac{DAT} and \ac{PIT}.
\ac{DAT} forces the model to learn to associate its output with the speaker order in the audio.
For the two-speaker case, the first separation encoder learns to focus on the leading speaker, and the second on the follow-up speaker if it exists.
\ac{DAT} computes RNN-T loss \(N\) times, as described in Eq. \ref{eq:dat-loss} and depicted on Fig. \ref{fig:mt-ms-rnnt}.
In contrast, \ac{PIT} allows the separation encoder to flexibly match either of the speaker, as long as it minimizes the loss function during training.
Therefore, \ac{PIT} carries out \(N^2\) RNN-T loss computations that are required to find an optimal permutation \(\pi\) of speaker \( n \) from the set of permutations \( \mathcal{P} \) (Eq. \ref{eq:pit-loss}).
\vspace{-0.01cm}
\begin{align}
\mathcal{L}_{DAT} &= - \sum_{n} \log P(\mathbf{y}_{n} | \mathbf{h}_n) \label{eq:dat-loss} \\
\mathcal{L}_{PIT} &=  \min \limits_{\pi \in \mathcal{P}} - \sum_{n} \log P(\textbf{y}_{n} | \mathbf{h}_{\pi(n)} ) \label{eq:pit-loss}
\end{align}

While \cite{SklyarPiunovaLiu_streaming_rnnt_icassp2021} showed that \ac{PIT} is better than \ac{DAT} at recognizing overlapping speech, \ac{PIT} generally has challenges in scaling up to a large number of speakers due to its \(O(N!)\) complexity.

\section{Technical approaches}
\label{sec:technical-approach}

\subsection{On-the-fly multi-speaker speech audio simulation}

Previous work \cite{kanda2021largescale} showed the benefit of pre-training on the artificial multi-speaker data simulated on-the-fly in comparison to the training on real multi-speaker data from scratch.
In this work we benchmark on-the-fly simulation against training on the fixed simulated data in close-talking conditions.
We demonstrate that due to the increased variety of overlapping speech scenarious seen during training it improves generalization to unseen multi-speaker audio.

To simulate the audio mixture, we select $S$ utterances randomly from a pool of single-speaker utterances.
For each follow-up utterance, we sample a random time delay from a specified delay range.
Before mixing, a reference speaker is randomly determined and the speech energy of other speakers is normalized to achieve an energy ratio randomly sampled from a specified range.
To simulate far-field recording scenario we convolve each source audio with an \ac{AIR} before padding with the sampled delay.

\subsection{Overlap-based target arrangement and \ac{MT-RNN-T}}
\label{subsec:multi_turn_target_assignment}
\vspace*{-2mm}
Along with the audio simulation of multi-speaker speech, the transcripts should also be combined on-the-fly accordingly.
For two-speaker \ac{MS-RNN-T} mixture, when the audio contains two overlapping speakers, the two reference transcripts are simply assigned to two targets.
When there is no overlapped speech, there are different strategies to assign the reference.
Previous work \cite{SklyarPiunovaLiu_streaming_rnnt_icassp2021} adopted speaker-based target arrangement (Fig. \ref{fig:mt-ms-rnnt} blue), which always assigns reference transcripts to different targets.
This strategy encourages the model to separate speaker by dedicating each output for one unique speaker, but it does not scale up to a large number of speakers in audio.
Here we propose a different strategy, overlap-based target arrangement, which allows to build two targets for two model outputs by combining references into two targets in the order of their appearance in the audio.
Overlap-based arrangement concatenates utterances together if there is no overlap between them and only incurs target switch at speech overlap between two consecutive utterances, as shown on Fig. \ref{fig:mt-ms-rnnt} (red).
This simple change in the training data pipeline allows us to train models on data with more speakers \(S\) and speaker turns \(U\)  than  the number of output channels \(N\): \(U \geq S > N\). Therefore, we refer to models trained with
overlap-based target arrangement as multi-turn RNN-T (MT-RNN-T).

We observed that overlap-based target switch is similar to the recent graph-PIT work \cite{Neumann21_graph-PIT}, though our work was completed fully in parallel and independently.
The target assignment is equivalent to graph-PIT but we do not permute different combinations of reference targets and model hypothesis for loss calculation. This effectively leads to complexity reduction from \(O(N^U)\) in graph-PIT to \(O(N)\) in this work.

\subsection{Rich transcriptions with speech segmentation}
\label{subsec:ritch_transcription}
\vspace*{-2mm}
Overlap-based target arrangement in \ac{MT-RNN-T} makes speaker tracking more difficult since multiple speakers can be assigned to the same output channel.
For downstream applications, speaker tracking in the multi-speaker \ac{ASR} system is important to provide not only ``what'' was said, but also ``who'' said what.
As the first step towards such rich transcriptions, in this work we introduce segmentation tags to indicate potential point of speaker turn change, so that future work on speaker labelling could be built on top.
To emphasize this conceptual difference from other research work based on speaker change detection \cite{k2020serialized,kanda2021slt}, we refer to each segment as a ``turn'', without necessarily insisting the semantic correctness of this term.
We augment model vocabulary with a ``change-of-turn'' (\texttt{<cot>}) token which is inserted between two consecutive turns in the same target, as depicted on Fig.  \ref{fig:mt-ms-rnnt}.

\section{Experiments}
\label{sec:experimental-results}

\subsection{Experiments on LibriSpeechMix}
\label{sec:experiments-on-librispeech-mix}
We perform our first set of experiments on LibriSpeechMix to verify the effectiveness of on-the-fly data simulation setup on a simpler task with limited number of speakers.

\begin{table}[t]
	\caption{Benchmarking on-the-fly-simulation with MS-RNN-T model variants against other streaming methods on LibriSpeechMix.}
	\label{tab:librispeech-mix-results}
	\centering
	\begin{tabular}{ l c c }
		\toprule
		Model & \makecell{On-the-fly \\ simulation}   & WER [\%] \\
		\midrule
		SURT \cite{Lu2021} & \xmark &  10.8 \\
		SURT \cite{lu2021streaming}  & \xmark  & 10.3 \\
		\midrule
		DAT-MS-RNN-T \cite{SklyarPiunovaLiu_streaming_rnnt_icassp2021} &  \xmark & 11.0  \\
		DAT-MS-RNN-T (this work)  &  \cmark &  9.2  \\
		PIT-MS-RNN-T \cite{SklyarPiunovaLiu_streaming_rnnt_icassp2021} &  \xmark & 10.2  \\
		PIT-MS-RNN-T (this work)  &  \cmark &  8.8  \\
		\bottomrule
		\vspace{-18pt}
	\end{tabular}

\end{table}

\vspace*{2mm}
\noindent
\textbf{Task description --}
LibriSpeechMix is a simulated dataset initially proposed in  \cite{k2020serialized} that contains artificially mixed utterances from LibriSpeech corpus \cite{Panayotov_libri}.
We use 2-speaker dev and test partitions of this dataset with a single turn for each speaker in the utterance. Delay for the second speaker is randomly sampled with the constraint that
each audio mixture has an overlapping segment. The original training dataset of LibriSpeechMix follows a similar simulation strategy with an additional constraint on a minimal delay of
0.5 sec between start times of speaker turns. It contains $\sim$1.5k hours of simulated data, and we substitute it with the proposed on-the-fly data simulation pipeline in the experiments
presented below.

\begin{table*}[t]
	\caption{\ac{MT-RNN-T} model benchmarking on LibriCSS with performance reported in optimal reference combination \ac{WER} [\%].}
	\label{tab:libricss-results}
	\centering
	\begin{tabular}{ c c c  c c c c c c c c}
		\toprule
		\makecell{Model\\ ID} & \makecell{Model\\ type} & \makecell{Max {\#}spk in \\ training} & \texttt{<cot>} & 0L & 0S & OV10 & OV20 & OV30 & OV40 & full \\
		\midrule
		0 &MS-RNN-T & 2  & \xmark & 16.1 & 15.8 & 25.0 & 35.2 & 46.0 & 48.0 & 32.8 \\
		\midrule
		1 &MT-RNN-T & 3  & \xmark  & 14.9  & 15.4 & 23.3  & 31.0  & 35.0  & 39.0 & 27.8  \\
		2 &MT-RNN-T & 4  & \xmark  & 15.1  & 15.3  & 20.6  & 27.2 & 34.0 & 36.8 &  26.0\\
		3 &MT-RNN-T & 5  & \xmark  & 14.8  & 14.5  & 18.0  & 25.8  & 30.3  & 32.3 & 23.6 \\
		\midrule
		4 &MT-RNN-T & 5  & \cmark  & 14.7  & 14.8  & 20.7  & 25.3  & 33.2  & 36.4 & 25.3 \\
		\bottomrule
	\end{tabular}
\end{table*}

\vspace*{2mm}
\noindent
\textbf{Training setup --}
The model topology of \ac{MS-RNN-T} is based on the one established in \cite{SklyarPiunovaLiu_streaming_rnnt_icassp2021}.
We use 2 LSTM layers in each recurrent module of the architecture (mixture encoder, 2 separation encoders, recognition encoder, prediction network) with 1024 units in each layer.
Output layers in the recognition encoder and the prediction network have 640 units. The joint network has a single feed-forward layer with 512 units.
The output softmax layer has dimensionality of 2501 which corresponds to the blank label and 2500  wordpieces that represent the most likely subword segmentation from
a unigram word piece model \cite{kudo2018sentencepiece}.
Acoustic features are 64-dimensional log-mel filterbanks with a frame shift of 10ms which are stacked and downsampled by a factor of 3.
We use SpecAugment with LibriFullAdapt policy \cite{Park_2020} for feature augmentation. We use the Adam algorithm \cite{kingma2014adam} with the warm-up, hold and decay schedule proposed in \cite{Park_2019} for the optimization of all models.
We select the best model checkpoint based on performance on the development set.

Similar to the previous work \cite{SklyarPiunovaLiu_streaming_rnnt_icassp2021} we pre-train MS-RNN-T model with a single separation encoder on the LibriSpeech dataset before training with both separation encoders on the simulated multi-speaker data. During on-the-fly simulation we randomly sample two utterances from different speakers in the pool of LibriSpeech data.
Consistent with the simulation configuration of the original LibriSpeechMix training corpus, time delay for the second speaker is sampled uniformly from the range: \( (0.5, \textrm{len}(utt_1)) \), while original signal energy is remained intact.

\vspace*{2mm}
\noindent
\textbf{Results --}
For the evaluation we used the same metric as in \cite{SklyarPiunovaLiu_streaming_rnnt_icassp2021}, i.e. \ac{OED WER}, that is also aligned with \cite{Tripathi_2020, k2020serialized}.
Table \ref{tab:librispeech-mix-results} shows the benefit of on-the-fly simulation on \ac{MS-RNN-T} models based on both \ac{DAT} and \ac{PIT}.
We compare their performance with the corresponding models trained on the fixed simulated dataset in \cite{SklyarPiunovaLiu_streaming_rnnt_icassp2021},
external \ac{SURT} proposed in \cite{Lu2021} and its improved variant from \cite{lu2021streaming}. All models in this comparison have streamable architectures and
comparable number of parameters ($\sim$81M).
On-the-fly simulation brings substantial improvements for both \ac{DAT} and \ac{PIT} training variants of \ac{MS-RNN-T} with 16\% and 14\% relative WER reductions, respectively.
It is noteworthy that relative WER difference between \ac{DAT} and \ac{PIT} training approaches diminish from 7\% to 4\%  when models are exposed to more diverse multi-speaker data generated on-the-fly.
\ac{PIT-MS-RNN-T} with on-the-fly-simulation achieves the new state-of-the-art performance on this benchmark among all streaming methods and outperforms the best external \ac{SURT}
model by 14\% relative WER.

\subsection{Experiments on LibriCSS}
\label{sec:experiments-on-libricss}
We perform our second set of experiments on LibriCSS dataset where we benchmark \ac{MT-RNN-T} model variants on partially overlapped multi-turn speech from more than 2 speakers.


\vspace*{2mm}
\noindent
\textbf{Task description --}
LibriCSS was originally proposed in \cite{chen2020continuous} for continuous speech separation.
It contains 10 far-field audio recordings with LibriSpeech utterances played back in a room to simulate meetings with 8 speakers.
The full LibriCSS evaluation data is divided into 6 partitions.
2 partitions exclude overlapped speech, with either short (0S) or long (0L) silence gaps between speaker turns.
The rest 4 partitions cover different overlap ratios, from 10\% to 40\%: OV10, OV20, OV30 and OV40.
We perform segmentation of the original one-hour long LibriCSS sessions into utterance group segments using oracle silence boundary information as described in \cite{kanda2021slt}.
It ensures the existence of both single-turn (single utterance) and multi-turn segments in the evaluation data.
This type of multi-speaker \ac{ASR} model benchmarking is also known as utterance group evaluation in the literature \cite{kanda2021largescale}, and it is important to diffentiate it
 from utterance-wise and continuous input evaluation protocols discussed in \cite{chen2020continuous}.
We use Session 0 of the dataset as a development set to tune the hyper-parameter word-reward \cite{RabinerJuang1993, HuangAcero2001} and to select the best checkpoint in a grid search fashion.
The remaining Sessions 1-9 are used to report performance.


\vspace*{2mm}
\noindent
\textbf{Training setup --}
We closely follow training setup used in LibriSpeechMix experiments  (Section \ref{sec:experiments-on-librispeech-mix}) for experiments on LibriCSS, with a few important differences.
First, we introduce layer normalization \cite{ba2016layer} in each LSTM layer of the model architecture, which we find
beneficial to improve convergence on more challenging training data.
Second, we extend on-the-fly simulation to support far-field conditions with multi-party speech.
We sample random number of utterances uniformly from the range $\{1, \dots, S \}$.
We do not impose the constraint on all utterances to come from different speakers, however the overlap between segments uttered by the same speaker is not allowed.
Each utterance \(s\) is scaled and convolved with \ac{AIR} before adding to the mixture.
Time delay for each subsequent overlapping utterance \(utt_{s+1}\) is sampled from the range  \( (0.5, \textrm{len}(utt_s)) \), which ensures overlapped speech segment existence in all simulated multi-speaker utterances.
Energy ratio between the reference speaker and each other speaker is sampled uniformly from the range -5 dB to 5 dB.
We filter out simulated audio if its length exceeds 30 seconds to avoid out-of-memory errors in the RNN-T loss.
This hardware limitation makes it inefficient to experiment with large maximum number of speakers \(S\) in the simulation, therefore we did not go beyond \(S=5\)
in the experiments.

\vspace*{2mm}
\noindent
\textbf{Evaluation metric --}
In our early experiments we found that the previously used \ac{OED WER} metric could not well represent the recognition accuracy on LibriCSS test data due to
the difficulty in predicting from which output thread the model will output the hypothesis for a specific region of audio.
Therefore, to provide an accurate picture of the recognition performance, we propose a new evaluation metric where the reference arrangement is optimized based on model output to provide the lowest WER readings, i.e. \ac{ORC WER}.
Assuming that there are $U$ reference transcripts for the audio, the number of transcripts that can be assigned to the first model output varies from $0$ to $U$, while the rest is assigned to the second output.
Then \ac{ORC WER} can be efficiently computed as follows.
First, we generate all $C_U^k$ possible reference combinations that assign reference transcripts to two model outputs, where $k \in \{0, \dots, U \}$ is the number of reference transcripts to assign to the first model output.
Then, we sort the references by start time and concatenate references within each generated combination.
For each combination, we compute the overall WER of two model outputs against two concatenated reference combinations, then report the lowest WER among all combinations.
This approach factors out the reference-hypothesis pairing errors from the actual word recognition errors.
In the evaluation, we left out the longest utterance (with 24 turns) obtained after segmentation from \ac{ORC WER} scoring as we found it computationally infeasible to search over all possible reference combinations.
In addition, segmentation token \texttt{<cot>} is excluded from hypotheses for scoring.





\vspace*{2mm}
\noindent
\textbf{Results --}
Evaluation results on LibriCSS dataset are presented in Table \ref{tab:libricss-results}.
We use \ac{MS-RNN-T} trained with on-the-fly simulated audio containing up to 2 speakers (model 0) as the baseline.
Since switching from \ac{MS-RNN-T} to \ac{MT-RNN-T} allows us to train on simulated data with more than 2 speakers,
in models 1-3 we incrementally increase the maximum number of speakers in training from 2 to 5. We observe that each incremental increase of this simulation parameter
 consistently improves the performance in all test-set partitions, with 15\%, 21\% 28\% relative improvements reported overall for each experimental model.
Comparing models 3 and 0, we observe a large reduction in deletion errors which is as high as 52\% relative.
Breaking down model performance by
regions with fully overlapped and non-overlapped speech, we observe that model 3 achieves larger gains on the latter (31\% relative WER improvement),
while performance gains on the fomer is slightly less pronounced (24.3\%). Model 3 still performs significantly worse on fully overlapped segments (50.6\% WER) than on non-overlapped ones (16.3\%), which reveals that separation in the encoder is far from ideal and can be improved.

\vspace*{2mm}
\noindent
\textbf{Turn counting accuracy --}
In this section we investigate transcription segmentation quality by comparing the number of turns estimated by the model with the real number of turns in the audio.
To do so, we firstly introduce \texttt{<cot>} token into the model vocabulary and target transcriptions as described in Section \ref{subsec:ritch_transcription}. Resulting
model is trained using the same audio simulation configuration as model 3, and is denoted as model 4 in Table \ref{tab:libricss-results}. Its speech recognition performance
is slightly worse than the corresponding model without rich transcription capability, which we attribute to the increased deletion rate by 30\% relative. We report its turn counting accuracy in Table \ref{tab:turn-counting-accuracy-libricss}. According to the results, model tends to consistently underestimate the number of turns.
Turn counting accuracy (in bold) drops below 50\% for utterances with more than 3 turns and it is around 0\% for utterances with more than 6 turns (omitted in the table).
We attribute it to the fact that the model was never exposed to audio with more than 5 turns.
This observation clearly identifies the room for potential improvement of the proposed speech segmentation strategy. Finding a better trade-off between under-segmentation and over-segmentation is of particular importance for downstream application of speaker labeling on top of \ac{MT-RNN-T}   output. One could argue that over-segmentation is preferable here since it is less detrimental and potentially recoverable as part of the speaker labeling pipeline.

\begin{table}[]
	\caption{Confusion matrix for number of turns estimated on LibriCSS by model 4: MT-RNN-T with \texttt{<cot>} token.}
	\label{tab:turn-counting-accuracy-libricss}
	\centering
	\setlength{\tabcolsep}{2pt}
	\begin{tabular}{c c c c c c c}
		\toprule
		\multirow{ 2}{*}{\makecell{Actual \\ \# turns}}  & \multicolumn{6}{c}{Estimated \# turns} \\
		&  1 & 2 & 3& 4& 5& 6 \\
		\midrule
		1 & \textbf{100.00} & 0.00 & 0.00 & 0.00 & 0.00 & 0.00  \\
		2 & 9.38 & \textbf{90.34} & 0.28 & 0.00 & 0.00 & 0.00 \\
		3  & 1.53 & 40.46 & \textbf{58.02} & 0.00 & 0.00 & 0.00  \\
		4  & 0.00 & 16.67 & 48.15 & \textbf{35.19} & 0.00 & 0.00  \\
		5 & 0.00 & 6.90& 29.31 & 58.62 & \textbf{5.17} & 0.00 \\
		6 & 0.00 & 0.00& 16.22 & 62.16 & 18.92& \textbf{2.70}  \\
		\midrule
		\vspace{-23pt}
	\end{tabular}
\end{table}

\section{Conclusions}
\label{ap:conclusion}

In this paper we successfully scaled up streaming multi-speaker RNN-T to audio with any number of speakers without compromising streaming ability or changes in the model architecture.
The on-the-fly data simulation system stands at the heart of the proposed work.
It can simulate speech audio with any number of speakers, both close-talking and far-field, with dynamic multi-turn target arrangement and speech energy variation.
On close-talking two-speaker LibriSpeechMix we achieved the new state-of-the-art 8.8\% WER among streaming model architectures.
We proposed overlap-based target arrangement in a \ac{MT-RNN-T} model that re-uses the same output channel when there is no speech overlap.
We trained this model on data with up to 5 speakers, and benchmarked it against two-speaker \ac{MS-RNN-T} model on LibriCSS test set.
We proposed \ac{ORC WER} for factored study on word recognition performance to address the challenges in scoring long form output from models with two parallel output channels.
In addition, we experimented with joint speech recognition and segmentation, as the first step to rich transcriptions for streaming speaker attributed ASR.
Through turn counting analysis of segmentation performance, we demonstrated that \ac{MT-RNN-T} models with turn segmentation tokens currently tend to underestimate the number of speech segments, which lays the foundation for potential future work.




\vfill
\pagebreak
%

\bibliographystyle{IEEEtran}
{\footnotesize
	\bibliography{strings,refs}
}
\end{document}